\documentclass{iau}
\usepackage{graphicx}
\usepackage{natbib}
\usepackage{amssymb,amsmath}

\title[Multiple origins of asteroid pairs] 
{Multiple origins of asteroid pairs}

\author[Seth A. Jacobson]   
{Seth A. Jacobson$^{1,2}$}

\affiliation{$^1$Laboratoire Lagrange, Observatoire de la C{\^o}te d'Azur, \\ 
Boulevard de lÕObservatoire, CS 3f4229, F 06304 Nice Cedex 4, France \\
$^2$Bayerisches Geoinstitut, Universt{\"a}t Bayreuth, \\ 
D 95440 Bayreuth, Germany \\
email: {\tt seth.jacobson@oca.eu}}

\pubyear{2015}
\volume{318}  
\pagerange{1}
\jname{Asteroids: New Observations, New Models}
\editors{S. Chesley, R. Jedicke, A. Morbidelli}
\begin{document}

\maketitle

\begin{abstract}
Rotationally fissioned asteroids produce unbound daughter asteroids that have very similar heliocentric orbits. Backward integration of their current heliocentric orbits provides an age of closest proximity that can be used to date the rotational fission event. Most asteroid pairs follow a predicted theoretical relationship between the primary spin period and the mass ratio of the two pair members that is a direct consequence of the YORP-induced rotational fission hypothesis. If the progenitor asteroid has strength, asteroid pairs may have high mass ratios with possibly fast rotating primaries. However, secondary fission leaves the originally predicted trend unaltered. We also describe the characteristics of pair members produced by four alternative routes from a rotational fission event to an asteroid pair. Unlike direct formation from the event itself, the age of closest proximity of these pairs cannot generally be used to date the rotational fission event since considerable time may have passed.

\end{abstract}
%
Asteroid pairs are a population of main belt asteroids that have nearly identical heliocentric orbits. These occur at a frequency in excess of that expected by random fluctuations of background asteroid orbit density~\citep{Vokrouhlicky:2008dt,Pravec:2009hv}. Furthermore, when the orbits of these asteroid pairs are carefully integrated backwards, many have close encounters in phase space (small distance and low velocity) in the recent past~\citep{Vokrouhlicky:2009kt}. This suggested a common origin, and the YORP-induced rotational fission hypothesis was immediately proposed~\citep{Vokrouhlicky:2008dt}. Matching spectral types between pair members complements a common origin hypothesis~\citep{Duddy:2012gb,Moskovitz:2012ga,Wolters:2014gn}. If a common origin hypothesis is accepted than these pairs provide powerful tools to test many theories including space weathering, mutual body tides and binary evolution. 
\section{YORP-induced rotational fission hypothesis}
The YORP-induced rotational fission hypothesis is a widely explored set of arguments that endeavors to explain the existence and properties of asteroid pairs, binaries and other multi-member systems from the rotational fission of rubble pile asteroids due to the rotational acceleration of the YORP effect~\citep{Rubincam:2000fg,BottkeJr:2002tu,Scheeres:2007io,Cuk:2007gr,Walsh:2008gk,Pravec:2010kt,Jacobson:2011eq}. The YORP effect is the spin and orbit averaged rotational torque on an asteroid due to thermal radiation~\citep{Rubincam:2000fg}, and its effects have been directly observed in nature~\citep{Vokrouhlicky:2003to,Taylor:2007kp,Lowry:2007by,Kaasalainen:2007hq}. The theory for this effect is well-developed for principal axis rotating asteroids~\citep[e.g.][]{Vokrouhlicky:2002cq,Breiter:2007gl,Nesvorny:2007gz,Scheeres:2007kv}, and the rotational acceleration is:
\begin{equation}
\dot{\omega}_Y = \frac{3 Y}{4 \pi \rho R^2} \left( \frac{G_1}{a_\odot^2 \sqrt{1-e_\odot^2}} \right)
\end{equation}
where $G_1 = 10^{14}$ kg km s$^{-2}$ is the solar constant at an astronomical unit divided by the speed of light, $a_\odot$ and $e_\odot$ are the heliocentric semi-major axis and eccentricity, respectively, $\rho$ and $R$ are the density and radius of the asteroid, respectively, and $Y$ is the YORP coefficient~\citep{Scheeres:2007kv}. The YORP coefficient is mostly dependent on the shape and orientation of the body but explicitly not size-dependent and observed values lay between 10$^{-3}$ and 10$^{-2}$~\citep{Taylor:2007kp,Kaasalainen:2007hq}. While the YORP effect is prevalent on all small bodies in the near-Earth region, the strong dependence of the rotational acceleration from the YORP effect on size and distance limits the consequences of this phenomena to small asteroids with radii $R \lesssim 6$ km in the main belt~\citep{Jacobson:2014bi}. However, there is a less-developed so-called tangential component that can increase the strength of the YORP torque for prograde rotators by a factor of two and so may extend this size domain by a few tens of percent; perhaps more importantly, this tangential component has a prograde bias~\citep{Golubov:2012kt,Golubov:2014hf}.

For asteroids undergoing spin up due to the YORP effect, centrifugal accelerations increase throughout the body, but they are resisted by gravity, cohesive forces and mechanical strength. Centrifugal accelerations match gravitational accelerations for a planar two sphere approximation, when the spin rate $\omega$ reaches a critical value:
\begin{equation}
\label{eqn:disruptionlimit}
\omega_c = \omega_d q_d \qquad q_d = \sqrt{\frac{1 - q^{1/3} + q^{2/3}}{\left(1+q^{1/3} \right)^2}}
\end{equation}
where $\omega_d = \sqrt{4 \pi G \rho / 3}$ is the critical disruption spin rate for a test particle on the surface of a constant density sphere, $G$ is the gravitational constant, $\rho$ is the density of the asteroid, $q_d$ is a function of the mass ratio $q$, which is the mass of the smaller, secondary component $m_s$ divided by the mass of the larger, primary component $m_p$~\citep{Hartmann:1967dt}. If the connection between the two components does not possess any strength, then when the spin rate reaches this critical value, the two components enter into orbit about each other.

If the internal structure of the asteroid has strength, then the asteroid goes into tension as the YORP effect increases the spin rate. For the planar two sphere approximation, the expression for the critical spin rate can be expressed simply as:
\begin{equation}
\label{eqn:strengthlimit}
\omega_c = \sqrt{ \left( \omega_d q_d \right)^2 + \left( \omega_t q_t \right)^2} \qquad q_t = \sqrt{\frac{ \left(1 - q^{1/3} + q^{2/3} \right) \left( 1 + q \right)^{2/3}}{q^{1/3}}}
\end{equation}
where $\omega_t = \sqrt{3 \sigma_c / \left( 4 \pi \rho R^2 \right)}$ is a simple prescription for the critical rotation needed to disrupt a progenitor asteroid of effective radius $R$ with a critical strength $\sigma_c$ and no gravity~\citep{Sanchez:2014ir}, and the factor $q_t$ scales the area of the stressed internal surface with the size of the smaller, secondary fissioned mass. Naturally, failure occurs along the weakest internal surface in the body and this determines the mass ratio $q$ of the orbiting components. Note, while this equation is dimensionally correct and acceptable for order of magnitude estimates, more detailed and accurate formulae are available that are derived directly from the application of the Drucker-Prager criterion on the von Mises stress and incorporate the role of cohesion as well~\citep{Holsapple:2001di,Sharma:2009fo,Sanchez:2014ir}. Furthermore, we note the similarity between our simple critical spin limit calculated above and those calculated by~\citet{Sanchez:2014ir}. They consider the role of cohesion as characterized by a material friction angle in detail and find that variations of the friction angle from 0$^\circ$ to 90$^\circ$ produce only a factor of two change in the critical tensile strength spin limit.  

There are only a few estimates of the critical strength of asteroids available. Examining the spin rate distribution of near-Earth and small main belt asteroids as a function of radius, there is a clear $\sim$2.3 hr spin period limit for asteroids larger than $\sim$0.2 km~\citep{Harris:1996tn,Pravec:2000dr,Pravec:2007ki}; some smaller asteroids are observed to rotate much faster~\citep{Pravec:2000dr,Hergenrother:2011jw}. This pattern and the lack of binaries with primary radii less than $200$ m can be understood if a typical asteroid critical strength is less than $\sim$25--100 Pa \citep{Sanchez:2014ir}. The rapidly rotating asteroid 29075 1950 DA requires a lower limit on the critical strength of $\gtrsim$75 Pa~\citep{Rozitis:2014bx,Hirabayashi:2015er}. The only direct estimate is derived from the disintegration of P/2013 R3 and suggests a critical strength of 40--210 Pa for this $165$ m radius asteroid~\citep{Hirabayashi:2014de}. For the ensuing simple calculations, an asteroid's critical strength is assumed to have an inverse dependence on the square root of the size of the body: $\sigma_c = \sigma_{c,0} \sqrt{ R_0 / R}$ where $\sigma_{c,0} = 125$ Pa and $R_0 = 165$ m. This relation follows the general understanding that mechanical strength is determined by flaws, which grow in size with the body, in analogy to Griffith's crack theory assuming a Weibull distribution of cracks.

\section{Free energy after a rotational fission event}
The critical spin rate determines the free energy of a newly formed binary system after a YORP-induced rotational fission event. The free energy is the sum of the rotational kinetic energy of each body, their relative translational kinetic energy, and their mutual gravitational potential energy. For two spheres in contact, which is the simplest approximation for a body undergoing rotational fission at the moment of fission~\citep{Scheeres:2007io}, the initial free energy $E_i$ of a rotationally fissioned binary is:
\begin{equation}
\frac{E_i}{E_c} =  q_1 q_d^2 - q_2 \qquad q_1 = \frac{2 + q \left( 7 + q^\frac{1}{3} \left( 10 + q^\frac{1}{3} \left( 7 + 2 q \right) \right) \right)}{q \left( 1 + q \right)^\frac{2}{3}} \qquad q_2 = \frac{10 \left(1 + q \right)^\frac{1}{3}}{1 + q^\frac{1}{3}}
\end{equation}
where $E_c = 2 \pi \rho q \omega_d^2 R^5 / 15 \left( 1 + q \right)^2$ is an energy normalization constant~\citep{Scheeres:2009dc}. 

\begin{figure}[t]
\begin{center}
\includegraphics{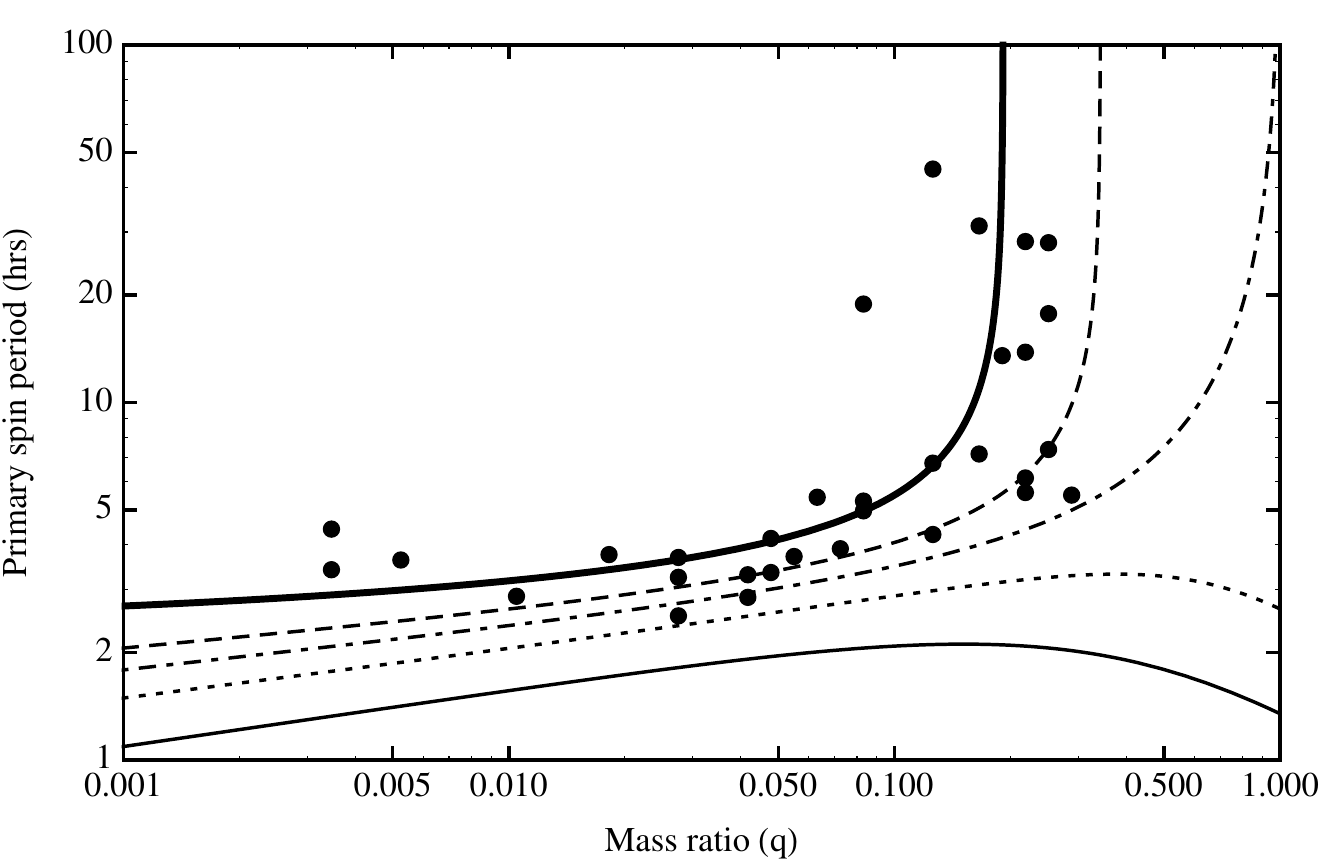} 
\caption{The final spin period of the larger, primary component of an asteroid pair after rotational fission as a function of mass ratio. The dots are measured spin periods and mass ratios of observed asteroid pairs from~\citet{Pravec:2010kt}. The lines show the spin period-mass ratio relationship for a planar, two sphere approximation for a rotationally fissioned progenitor with different strengths: strengthless ($\omega_t = 0$; bold, solid), while the others are $\omega_t = 0.243 \omega_d$ (dashed), $\omega_t = 0.324 \omega_d$ (dot-dashed), $\omega_t = 0.432 \omega_d$ (dotted), and $\omega_t = 0.648 \omega_d$ (thin, solid). The critical tensile strength spin rate is $\omega_t = \sqrt{3 \sigma_c / \left( 4 \pi \rho R^2 \right)}$, where the absolute tensile strength is $\sigma_c = \sigma_{c,0} \sqrt{ R_0 / R}$. With the caveat that these parameters ($\sigma_{c,0} = 125$ Pa and $R_0 = 165$) are poorly characterized as discussed in the text, the lines representing bodies with strength correspond to a range of sizes: 500 m, 400 m, 320 m and 230 m, respectively.}
\label{fig:WithStrength}
\end{center}
\end{figure}

This free energy is the energy accessible to each of the energy reservoirs: spin of each body and their relative motion. The spin-orbit coupling of the higher-order non-Keplerian gravitational potentials of the two components with orbit (and, to a lesser extent, tides) transfers energy between these reservoirs. If the energy in the orbital motion ever exceeds that in the gravitational potential then the system is on a disruption trajectory and will become unbound once the two components reach their mutual Hill sphere. This can only occur in the spherical approximation for asteroids with mass ratios $q \lesssim 0.2$, because only for these systems is the critical spin rate high enough that the free energy is positive and disruption trajectories are available~\citep{Scheeres:2007io}. As shown in Figure~\ref{fig:WithStrength}, the observed asteroid pairs obey the relationship between the rotation rate of the larger, primary member of the pair and the mass ratio predicted by considerations of the free energy of a YORP-induced rotational fission event, this is clearly the primary mechanism for the formation of asteroid pairs~\citep{Pravec:2010kt}. We briefly review the theory here and then expand the theory to consider the roles of strength and secondary fission on the relationship between mass ratio and primary spin rate. These may be the principal mechanisms for the production of outliers to the theory as presented in~\citet{Pravec:2010kt}.
\subsection{Asteroid pair formation directly from a strengthless asteroid}
Within the closed system of a binary formed after YORP-induced rotational fission, the free energy must be conserved. If the two components scatter each other onto a disruption trajectory due to higher-order non-Keplerian gravitational terms, i.e. spin-orbit coupling, then the free energy available to the spin states of each component is reduced since some energy is permanently stored in the disruption orbit. Also, the ratio of the rotational energy in the secondary to the primary is $\propto q^{5/3} \omega_s / \omega_p$, which means for similar spin rates and low mass ratios $q \lesssim 0.2$, the free energy stored in the secondary is relatively insignificant, so we assert as a first-order approximation that the secondary is rotating at the critical spin rate. Again, we use a planar two-sphere approximation, however it must be recognized that some asphericity is necessary for the two components to have found a disruption trajectory from their initially binary state after rotational fission, although the level of asphericity needed is very low for the binary system to explore phase space and find an escape trajectory since the initial semi-major axis of the system is very small compared to the radius of the primary~\citep{Jacobson:2011eq}. Then, given these assumptions, the free energy of the disrupted system is only the energy in the two spin states:
\begin{equation}
\frac{E_d}{E_c} = q_3 \left( q_d^2 + \frac{\omega_p^2}{\omega_d^2} q^{-5/3} \right) \qquad q_3 = 2 \left( q^2 \left( 1 + q \right) \right)^{1/3}
\end{equation}
where $\omega_p$ is the spin rate of the primary. Setting this energy equal to the free energy available after fission, the final spin rate of the primary is:
$\omega_p / \omega_d = q^{5/3} \left( \left( q_1 - q_3  \right) q_d^2 -q_2 \right) / q_3 $. This spin rate as a rotation period is plotted (bold, solid line) as a function of mass ratio in Figure~\ref{fig:WithStrength}. 

This simple relationship provides an easy to test hypothesis for the asteroid pair population, and the data fit the prediction well~\citep{Pravec:2010kt}, as shown. Some of the systems stray from the predicted relation, and \citet{Pravec:2010kt} explored how changes in the spherical approximation may account for these discrepancies. In particular, asserting that the components have different shapes changes the initial free energy system since the critical spin rate for rotational fission is a function of the distance between the mass centers of the components. We can imagine at least two other modifications to the scenario above that may create asteroid pairs that do not follow the above relation. 
\subsection{Asteroid pair formation directly from an asteroid with strength}
As discussed above, asteroids likely have significant tensile strength as their sizes decrease. If this is the case, then the free energy $E_i$ available to the binary system will be greater than in the case of only gravity, and using the same approximations as above for the free energy of the disrupted asteroid pair system, the final spin rate of the primary is:
\begin{equation}
\frac{E_i}{E_c} = q_1 \left( q_d^2 + \frac{\omega_t^2}{\omega_d^2} q_t^2 \right) - q_2 \qquad \frac{\omega_p}{\omega_d} = \sqrt{ \frac{ q^{5/3} \left( \left( q_1 - q_3  \right) \left( q_d^2 + q_t^2 \left(\frac{\omega_t}{\omega_d}\right)^2 \right) - q_2 \right)}{ q_3} }
\end{equation}
where the strength of the progenitor asteroid is characterized by the ratio of the critical tensile strength spin limit to the critical gravitational spin rate $\omega_t / \omega_d$. When this ratio is zero, the strengthless solution is obtained, but as this ratio increases, the primary spin rate after disruption increases for a given mass ratio. Because the spin rate necessary to fission the progenitor increases with $\omega_t$, increasingly higher mass ratio systems have positive free energies and can disrupt. For instance, the minimum necessary strength of the progenitor asteroid in order for it to spin fission in half (i.e. $q = 1$) is $\omega_t / \omega_d \approx 0.324$. Given the nominal asteroid strength relationship between critical tensile strength spin limit and size as determined above, this occurs for an asteroid with a radius of 400 m. A larger asteroid is unlikely to fission in half and form an unbound asteroid pair, while a smaller asteroid may fission in half and even have a rapidly rotating primary. In general, asteroids may span a wide variety of critical strengths due to variations in size and strength laws, and so a number of different $\omega_t / \omega_d$ ratios are plotted in Figure~\ref{fig:WithStrength}. 

In summary, if the progenitor asteroid has strength, then it is possible for the primary to rotate at low periods even when the mass ratio is relatively high. Moreover, the critical mass ratio is no longer $0.2$, but a function of the ratio $\left(\omega_t / \omega_d \right)$, which is directly related to the critical strength and therefore, the size of the progenitor asteroid.
\begin{figure}[t]
\begin{center}
\includegraphics{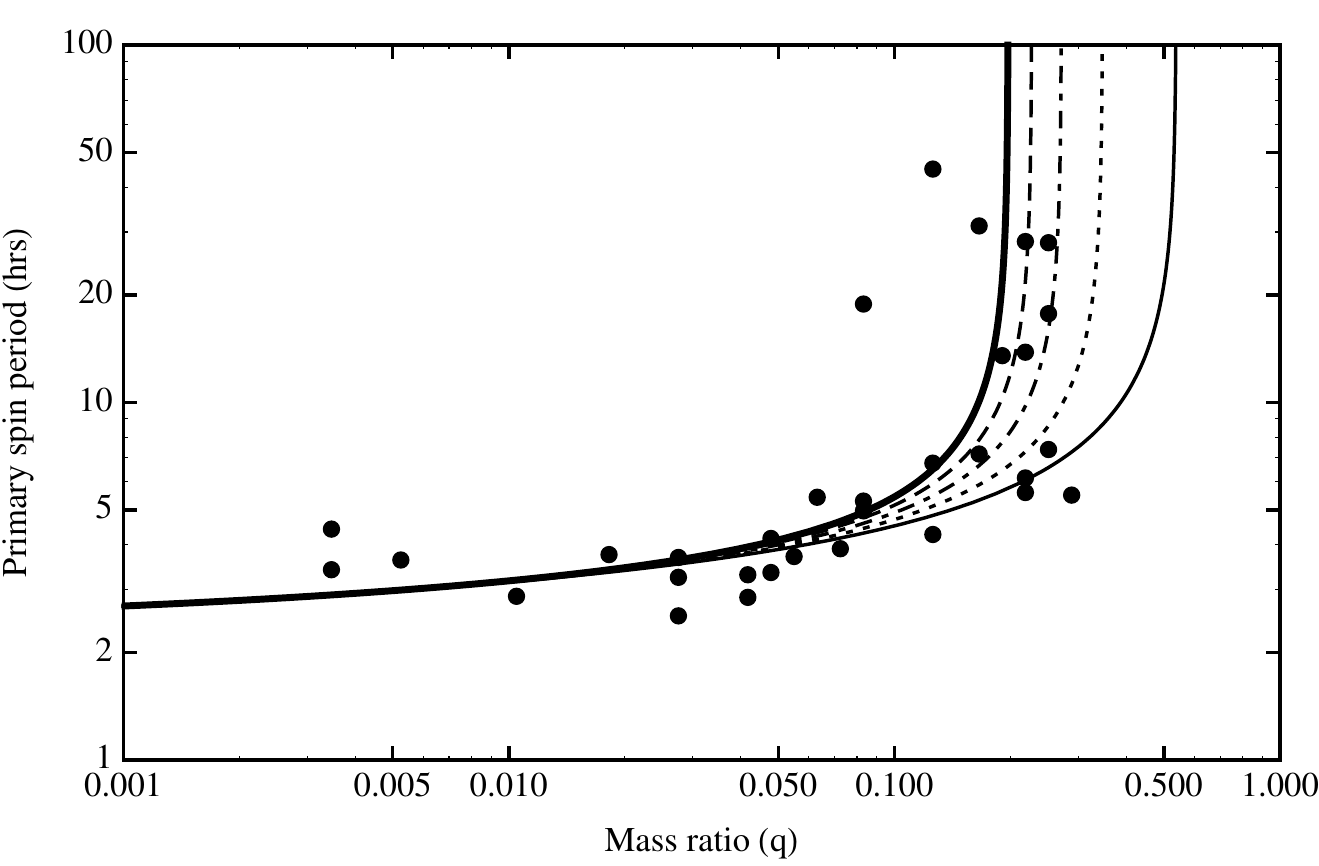} 
\caption{Similar to Figure~\ref{fig:WithStrength}, the final spin period of the larger, primary component of an asteroid pair after both rotational fission and secondary fission as a function of mass ratio $q$. The dots are measured spin periods and mass ratios of observed asteroid pairs from~\citet{Pravec:2010kt}. The lines show a planar, three sphere approximation for a paired binary system with semi-major axes of infinite (bold, solid), 6 (dashed), 3 (dot-dashed), 2 (dotted) and 1.5 (thin, solid) primary radii assuming an equal mass secondary fission event, so a mass ratio $p  = 1$.}
\label{fig:BinaryRole}
\end{center}
\end{figure}   
\subsection{Asteroid pair formation after a secondary fission event}
The asteroid pair production process was found to be incredibly efficient, and secondary fission was proposed as mechanism to stabilize rotationally fissioned asteroids to create long-lasting binary systems~\citep{Jacobson:2011eq}. Secondary fission occurs when the smaller, secondary binary member is torqued due to spin-orbit couping to a critical spin rate and thus, the secondary itself undergoes rotational fission. For simplicity, we only consider strengthless asteroids, so the initial free energy is the same as in that case. Now however, there is a binary system paired with the disrupted tertiary member, so the free energy includes terms for the spin states of all three bodies, their relative translational energies, and the mutual potential energy of the binary:
\begin{equation}
\frac{E_b}{E_c} = q_3 \left( p_1 + \frac{\omega_p^2}{\omega_d^2} q^{-5/3} \right) - q_2 \frac{1+q^{1/3}}{2 a \left( 1 + p \right) } \qquad p_1 = \frac{1 - \left(1 - p^{1/3} \right) \left(1 + p^{2/3} \right) p^{1/3} }{\left(1 + p^{1/3} \right)^2 \left( 1 + p \right)^{2/3} }
\end{equation}
where $p$ is the mass ratio of the smaller, tertiary component to the larger, secondary component after secondary fission. The final spin rate of the primary is then a function of both relevant mass ratios $q$ and $p$ as well as the semi-major axis $a$ measured in primary radii $R_p$ of the bound system:
\begin{equation}
\frac{\omega_p}{\omega_d} = \sqrt{ \frac{ q^{5/3} \left( \left( 1 + q^{1/3}  \right) q_2 - 2 \left(1 + p \right) \left( q_2 - q_1 q_d^2 + q_3 p_1 \right) a \right)}{ 2 q_3 \left( 1 + p \right) a} }
\end{equation}
This relationship between primary spin rate and the mass ratio of the bound system to the unbound pair member is shown in Figure~\ref{fig:BinaryRole}. While there is a dependence on the semi-major axis of the bound system, it's remarkable how unchanged the expected spin period of the primary in a paired binary compared to a lone primary. This is important, because the secondary fission hypothesis suggests that most low mass ratio binaries formed via this process~\citep{Jacobson:2011eq}. Indeed, two such systems have already been discovered: 3749 Balam is a triple system with an associated pair member~\citep{Vokrouhlicky:2009ja} and 8306 Shoko is a binary system with an associated pair member~\citep{Pravec:2013ud}. The creation of these binary systems simultaneously with their pair members means that binary evolution can be tied to a well estimated timescale. Thus, this enables a powerful technique to learn about tidal parameters and asteroid geophysics.
\begin{figure}[t]
\begin{center}
\includegraphics[width=\textwidth]{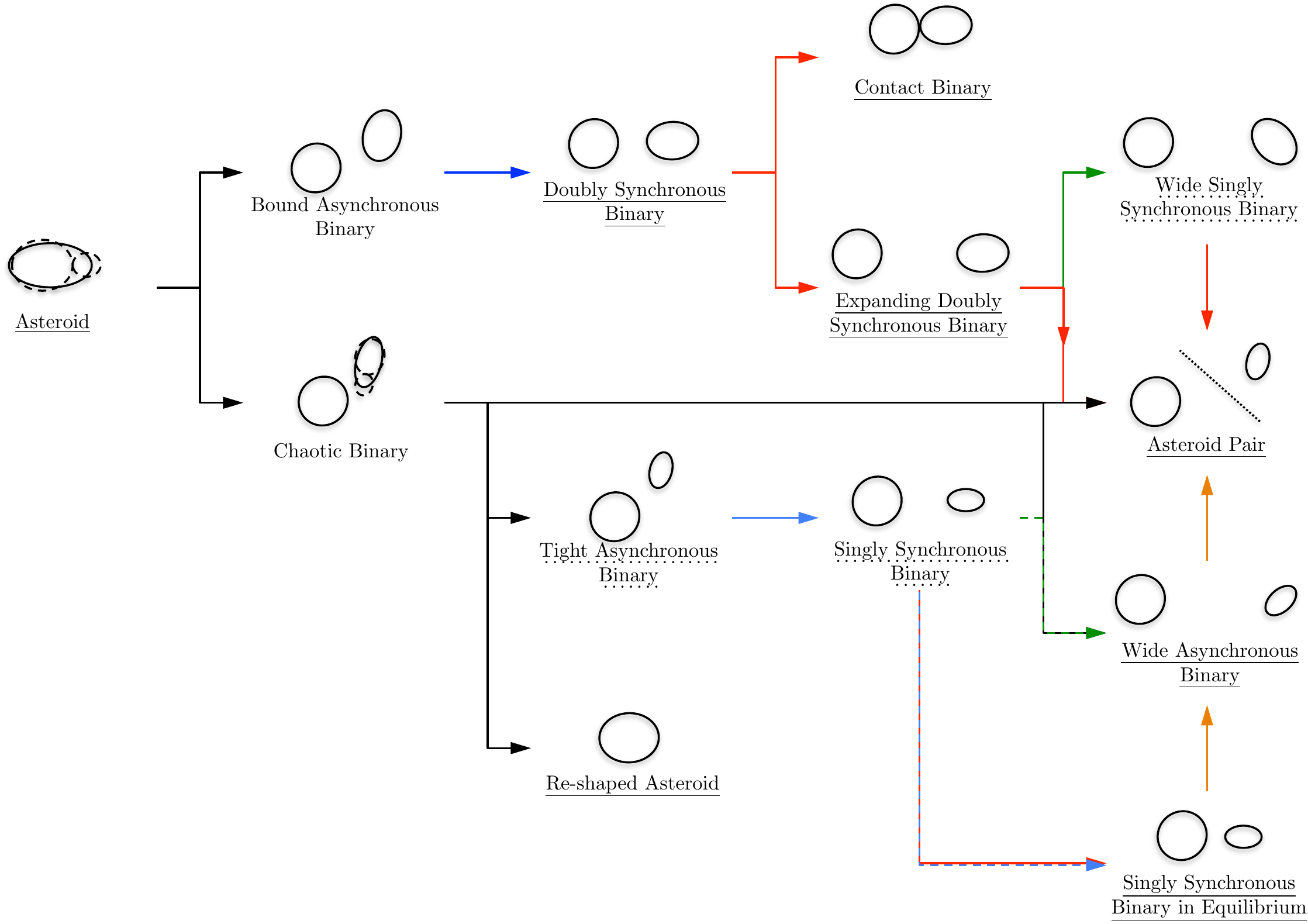} 
\caption{Flowchart depicting the evolution of an asteroid after a YORP-induced rotational fission event (reproduced from~\citet{Jacobson:2014jw}). Arrows indicate direction of evolution between cartoons and labels of the different evolutionary states. Solid and dashed underlines indicate long- and short-term stability, respectively. Colors indicate the dominant evolutionary process: black (purely dynamical), blue (tidal), red (BYORP effect), green (YORP effect), and yellow (planetary flyby).}
\label{fig:Flowchart}
\end{center}
\end{figure}   
\section{Alternative asteroid pair formation mechanisms}
The binary evolution model that has been developed to tie together the observed binary systems is reviewed in detail in \citet{Jacobson:2014jw}, and it is best summarized by the flowchart in Figure~\ref{fig:Flowchart}. If after YORP-induced rotational fission the asteroid is bound (negative free energy), then it follows the upper path, whereas if the asteroid is unbound (positive free energy) it follows the lower path. Asteroid pairs formed directly from rotational fission or secondary fission events are indicated by the middle path. This schematic shows four other paths to asteroid pair formation, which we discuss in turn below after introducing mutual body tides and the BYORP effect.

Mutual body tides are gravitational torques that arise from the delayed response of each binary component to the changing gravitational field of the system. If the bodies were inviscid fluids that reacted instantly to gravity, then these tides would be non-existent. However, real asteroids react viscously to tides dissipating energy in the form of heat and tidally torquing the orbit~\citep{Efroimsky:2015ia}. For the binary systems created by YORP-induced rotational fission, these tidal torques typically expand the semi-major axis and damp the eccentricity of the mutual orbit~\citep{Goldreich:2009ii}.

The binary YORP (BYORP) effect is similar to the YORP effect, but instead of an averaged thermal radiation torque on the spin state, it is an average thermal radiation torque on the mutual orbit. Typically this effect averages to zero, since the relative orientation of the binary components is effectively random with respect to the mutual orbit, but when either or both of the binary components is in a spin-orbit resonance, then the BYORP effect becomes non-negligible~\citep{Cuk:2005hb}. The direction of this torque on the orbit is dependent on the shape and orientation of the bodies in the spin-orbit resonance, and so it can either contract or expand the orbit~\citep{McMahon:2010by}.  

\subsection{Asteroid pair formation from singly synchronous mutual orbit expansion}
Binary asteroids formed from YORP-induced rotational fission with low mass ratios typically tidally synchronize only the secondary since the tidal synchronization timescales are so different~\citep{Goldreich:2009ii,Jacobson:2011eq}. Once synchronized, the secondary begins to evolved due to the BYORP effect~\citep{Cuk:2007gr,McMahon:2010jy}. If it evolves inward, it likely reaches a tidal-BYORP equilibrium, which matches the properties of most observed small binary asteroid systems~\citep{Jacobson:2011hp,Scheirich:2015ez}. However, if the BYORP effect expands the mutual orbit, then it may expand to the Hill radius. Given estimates of the strength of the BYORP effect~\citep{McMahon:2010jy,McMahon:2012ti}, semi-major axis growth may take $\sim$10$^5$--10$^7$ years before it reaches the Hill radius of the system.  Once it reaches the Hill radius, the binary will disrupt. 

Asteroid pairs created this route take significantly longer to form after the rotational fission event than asteroid pairs created directly from rotational fission, which may be formed only $\sim$0--10$^3$ years after the rotational fission event. This has significant implications for interpreting the dynamical ages of asteroid pairs determined from backwards integration of the heliocentric orbits. Furthermore, the secondaries of these asteroid pairs have been tidally decelerated before the system is disrupted. This is very different than the secondaries of asteroid pairs formed directly from rotational fission are as likely to still be rapidly rotating as not since the spin-orbit coupling during the temporary binary phase after YORP-induced rotational fission is best approximated by a random walk~\citep{Jacobson:2011eq}.

\subsection{Asteroid pair formation from doubly synchronous mutual orbit expansion}
Binary asteroids formed from YORP-induced rotational fission with negative free energy typically have high mass ratios and so tidally synchronize both binary components. Tidal synchronization timescales for small ($R <10$ km) asteroids in this configuration are typically less than or similar to near-Earth asteroid lifetimes and so significantly shorter than main belt collisional lifetimes~\citep{Goldreich:2009ii,Jacobson:2011eq}. Once tidally locked, these binaries undergo BYORP effect driven evolution. The direction of this mutual orbit evolution is dependent on the shapes and orientations of both binary members; it's possible that the torque on each member add in the same direction or subtract from each other, but it's unlikely that they perfectly cancel. Like above, semi-major axis growth may take $\sim$10$^5$--10$^7$ years before the mutual orbit expands to the Hill radius of the system. Once it reaches the Hill radius, the binary disrupts. 

Asteroid pairs created via this mechanism will have very different characteristics than those formed directly from YORP-induced rotational fission or after a secondary fission event. These asteroid pairs are likely to have high mass ratios and low spin rates. Furthermore, the dynamical age determined from backwards integrating the heliocentric orbits of the pair members, again like above, does not simply correspond with the date of the rotational fission events.
\subsection{Asteroid pair formation from mutual orbit expansion after de-synchronization of one member}
The most complicated scenario envisioned for the formation of asteroid pairs occurs when both members of doubly synchronous binary asteroids have expansive BYORP torques and that during this expansion, once of the binary components desynchronizes and begins to circulate due to the YORP effect. This is very similar to the process that forms the wide asynchronous binary population~\citep{Jacobson:2014hp}, but one of the components is still synchronous. Thus the BYORP effect continues to expand the system to the Hill radius and once it reaches the Hill radius, the binary disrupts. 

Asteroid pairs created by this chain of events would appear strange. They would have a high mass ratio, but one could be rotating quite rapidly (it could be either the primary or the secondary) while the other rotates very slowly. Like the previous two mechanisms, the dynamical age determined from backwards integration of the asteroid pair is not indicative of the timing of the rotational fission event.
\subsection{Asteroid pair formation from planetary flybys}
The simplest formation mechanism for asteroid pairs can only occur when binary asteroids are on planet crossing orbits. Then it is quite possible for planetary flybys to disrupt these binaries~\citep{Fang:2012go}, however the perturbations from the very planet that caused the disruption may make the determination of an asteroid pair quite difficult. This process could occur to any binary asteroid system morphology, so the observed asteroid pair properties would span a large range.
\section{Conclusions}
In this proceedings, we have summarized the evidence that most asteroid pairs form either directly from YORP-induced rotational fission or subsequent secondary fission events. This is important because asteroid pair formation is then intimately tied to binary formation. Indeed, the discovery that some binary asteroids like 8306 Shoko and 3749 Balam have unbound pair members is likely to create stringent constraints on binary evolution in the near future. We have also summarized the different asteroid pair formation mechanisms due to the BYORP effect and planetary flybys. These asteroid pairs can appear very different but also similarly to those that formed directly following a rotational fission event. Importantly, the interpretation of the dynamical age of the asteroid pair system is very different.

\bibliography{biblio}
\bibliographystyle{apalike}

\end{document}